\documentclass{aa}
\usepackage{graphics}

\begin{document}

   \thesaurus{11(11.13.1)} 

   \title{The morphology of the Magellanic Clouds revealed by stars of
different age: results from the DENIS survey}

   \author{Maria-Rosa L. Cioni
           \and	
           Harm J. Habing
           \and
           Frank P. Israel
          }

   \offprints{mrcioni@strw.leidenuniv.nl}

   \institute{Sterrenwacht Leiden, Postbus 9513,
              2300 RA Leiden, The Netherlands\\
             }

   \date{Received 20 April 2000 / Accepted}

   \titlerunning{The morphology of the Magellanic Clouds from DENIS}
   \authorrunning{Maria-Rosa L. Cioni et al.}

   \maketitle

   \begin{abstract}  The spatial  distribution  of sources  populating
   different  regions  of the  colour--magnitude  diagram ($I-J$, $I$)
   extracted from  the DENIS  catalogue towards the  Magellanic Clouds
   (DCMC  -- Cioni et  al.~\cite{cio}) reveal  significantly different
   morphologies. Each region is associated to a different age group.
    The  Large Magellanic Cloud (LMC)  shows an extended
   circular  shape with  a  prominent, off center  bar,  a nucleus  and
   irregular  spiral  arms.   The   Small  Magellanic  Cloud  shows  a
   perturbated  structure with  a prominent  central  concentration of
   stars. Old and young populations are offset from one another.

   \end{abstract}

\section{Introduction}
The  Magellanic  Clouds  are  our closest  neighbors  allowing  direct
observation of individual constituent  objects.  They are bound to the
Galaxy and show  signs of strong interaction with  the Milky Way about
$0.2$ Gyr  ago (Westerlund~\cite{west}).  The LMC is  classified as an
irregular dwarf galaxy,  its most prominent feature is  a central bar,
much like those  found in barred spiral galaxies.  Its eastern side is
closer  than  its  western   side  (Caldwell  \&  Coulson  \cite{cc}).
Underlying   the   bar   is   a   circular   disk   of   older   stars
(Westerlund~\cite{west}).  The appearance  of the SMC is characterized
by a  much less  pronounced bar, and  an eastern extension  called the
Wing.   Lines-of-sight  through  the  SMC appear  to  cover  extensive
depths; the Wing and the northeastern  part of the Bar are closer than
the southern parts (Westerlund \cite{west}).

Newly obtained  large photometric  data sets at  different wavelengths
and  with  improved  sensitivity  and  spatial coverage  allow  us  to
investigate the  large scale properties  of the Magellanic  Clouds. In
particular, data  in the  near infrared allow  us to access  stages of
stellar evolution that are marginally covered by optical data, such as
the RGB and AGB phases.

Very recently  Zaritsky et al.~(\cite{zar}) found  that the asymmetric
 appearance  of the  SMC is  primarily caused  by the  distribution of
 young  stars,  and   that  the  older  stars  have   a  very  regular
 distribution. It is  not possible from their Figures  to evaluate the
 behavior of the density towards the center of the Cloud.  Weinberg \&
 Nikolaev  (\cite{wn}) point  out  the presence  of intervening  tidal
 debris up to $\approx 15$ kpc from the LMC.
 
\section{The Data}
Our morphological study  of the Magellanic Cloud is  based on a sample
of   stars    extracted   from   the   DCMC    catalogue   (Cioni   et
al.~\cite{cio}). The sample includes  all sources detected in both $I$
and  $J$, irrespective  of  detection in  $K_S$.   The DCMC  catalogue
contains sources  detected in  at least two  of the three  DENIS bands
($I$:  $0.8\mu$m,  $J$: $1.25\mu$m  and  $K_S$:  $2.15\mu$m) within  a
surface   area   of  $20\times   16$   square   degrees  centered   on
$(\alpha,\delta)=(5^h27^m20^s$,$-69\degr00\arcmin00\arcsec)$     toward
the   LMC    and   $15\times   10$   square    degrees   centered   on
$(\alpha,\delta)=(1^h02^m40^s$,$-73\degr00\arcmin00\arcsec)$     toward
the  SMC;  J2000 coordinates  are  used  throughout  this paper.   The
observations have  been performed with the  DENIS instrument (Epchtein
et al.~\cite{epal}) on the 1m--ESO telescope.

\begin{figure}
\resizebox{6cm}{!}{\includegraphics{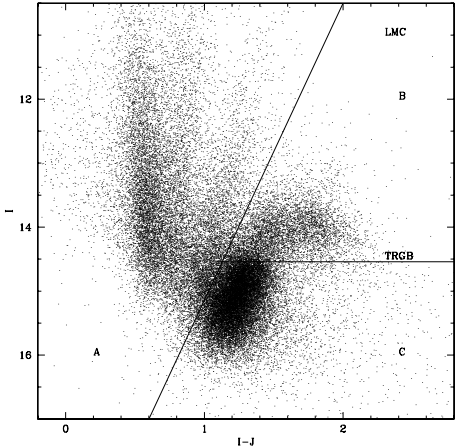}}
\caption{Colour--Magnitude  Diagram  ($I-J$,$I$)  of  sources  detected
simultaneously in $I$, $J$  and $K_S$ towards the LMC ($-69<\delta<-67$).
The horizontal  line marking the position of the  TRGB 
(Cioni  et al.~\cite{mr}), and the  slanted line at  
$I<-4.64\times(I-J)+19.78$  define the classes  A, B, and C discussed
in the text.}
\label{cml}
\end{figure}

We  have  used the  ($I-J$,  $I$)  colour--magnitude diagrams  (Figure
\ref{cml} for the LMC only) to select three classes of objects in each
Cloud. Sources labelled  (A) with $I<-4.64\times(I-J)+19.78$ represent
the youngest population in  the Magellanic Clouds: the brightest dwarf
stars, blue--loop stars and  supergiants (third vertical sequence from
the left),  together with an unrelated foreground  component of dwarfs
and  giants (first  two vertical  sequences from  the  left).  Sources
labelled (B) with $I>-4.64\times(I-J)+19.78$, located above the tip of
the  red giant  branch (TRGB  --  Cioni et  al.~\cite{mr}) are  mainly
asymptotic  giant  branch  stars  (AGB).  Sources  labelled  (C)  with
$I>-4.64\times(I-J)+19.78$ located below the TRGB are mostly red giant
branch stars (RGB) and represent  the oldest population in the Clouds.
For the sake of clarity,  we have plotted in Fig.~\ref{cml} only those
sources that were  detected in all three wave bands  and that occur in
the very central  part of the Cloud. Sources detected  only in $I$ and
$J$ predominantly populate the lower part of the diagram. The position
of the TRGB is indicated by  a horizontal line. The $I$, $J$ and $K_S$
sensitivity     limits    are    $18$,     $16$    and     $14$    mag
respectively. Photometric  errors widen the  sequences towards fainter
magnitudes.

For each  class of objects  in each of  the two Clouds, we  show their
distribution in the  plane of the sky by counting  the sources in bins
of  $0.2\degr \times  0.2\degr$,  applying a  light  smoothing to  the
resulting structure (Figs.~\ref{lmca}--\ref{smcc}); the contour values
increase  logarithmically.   Regions  corresponding  to  missing  data
(strips  at constant  RA  indicated  by diamonds)  were  filled in  by
interpolation.   Their   effect  is   mostly   negligible  except   in
Fig.~\ref{lmcc} where strips of possibly lower photometric quality may
be causing discontinuities in the outermost contours.

\section{Spatial distribution}

The  contribution  due  to  Galactic  foreground stars  has  not  been
subtracted  from the  maps.  Its  influence  is most  clearly seen  in
Fig.~\ref{lmca} in the  direction of the Galactic Plane.  In the other
maps,  the foreground  contribution is  rather constant  and  does not
affect the morphology of the Clouds.

\subsection{Structure of the LMC}

\begin{figure}
\resizebox{7cm}{!}{\includegraphics{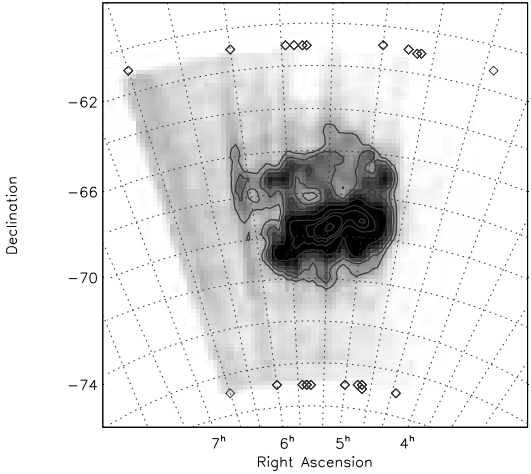}}
\caption{Star count of the LMC -- class A (Fig.~\ref{cml}); contours are at: 100, 125, 150, 200, 300, 400, 500, 800 per $0.04$ deg$^2$.}
\label{lmca}
\end{figure}

The  lower contours  in  Figs.~2--4 show  an  almost circular  outline
(axial ratios consistent with an  inclination $i = 30\degr - 40\degr$)
centered  in all  three cases  near  $\alpha =  5^{h}20^{m}, \delta  =
-69\degr$ with major axis at about $13^{\circ}$. Westerlund  (1997) gives a
similar diameter  for the  stars of the  old disk.  This  stellar disk
also  coincides  in  shape  and  extent  with  the  HI  disk  (Kim  et
al. 1998). The center of the disk is offset from the center of the Bar
by about 30$\prime$ to the north  (see Fig. 2).  We confirm the conclusion
by Westerlund (1997) that the  LMC consists of two systems: a circular
disk and an off  center bar. Half of the total number  of stars are in
the  bar  and  this  factor (Fig.~\ref{lmca})  increases  for  younger
objects.  Unless this is a transient configuration, it thus seems that
the LMC must  be embedded in a gravitational  potential produced by an
unseen mass component (see also Sofue 1999). This is in agreement with
the  conclusion by Stil  (1999) that  the class  of dwarf  galaxies to
which the LMC belongs (`fast rotators') is dominated by dark matter.

The youngest  component (younger than  $0.5$ Gyr) is composed  of very
bright  main-sequence dwarf stars,  blue--loop stars  and supergiants.
Their  distribution (Fig.~\ref{lmca})  is clumpy  and  irregular.  The
Bar, extending over  about $4\degr$, is prominent and  contains a well
defined nuclear concentration  at its center. The region  of 30~Dor is
represented by the  small feature just above the  northeastern side of
the Bar, and  the Shapley Constellation III is  the large structure at
$\delta  \approx  -67\degr$. Elongations  at  either  end  of the  Bar
indicate  the  presence  of  spiral  arms most  clearly  seen  in  the
northwest  at the  location  of  the giant  HII  region complex  N~11.
Similar  structures   are  seen  in  the   distribution  of stellar 
complexes (Maragoudaki et al.~\cite{mara}),  
associations and HII regions (Bica et al.~\cite{bical}). Clusters 
(Bica et al.~\cite{bisch}, Kontizas et al.~\cite{kon}) 
have a distribution more similar to the one of AGB/RGB stars.

\begin{figure}
\resizebox{7cm}{!}{\includegraphics{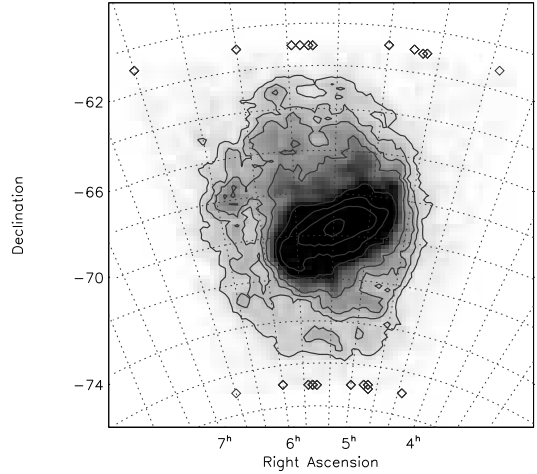}}
\caption{Star count of the LMC -- class B (Fig.~\ref{cml}); contours are at: 3, 5, 7, 10, 20, 30, 50, 100, 150 per $0.04$ deg$^2$.}
\label{lmcb}
\end{figure}

The distribution of AGB stars (Fig.~\ref{lmcb}), also relatively young
(around $1$ Gyr) likewise reveals a prominent Bar and nucleus. Shapley
Constellation III  is inconspicuous in  AGB stars.  A broad  and faint
spiral arm  begins at the northwestern  end of the  Bar and bifurcates
around $\alpha  = 05^{h}10^{m}, \delta = -66.5\degr$.   The spiral arm
feature  originating at  the southeastern  end of  the Bar  is clearly
delineated in  the AGB star population  and can easily  be followed to
$\alpha  = 4^{h}45^{m},  \delta =  -73\degr$. It  was noted  before by
Bothun \&  Thompson (\cite{bt}) in  their surface photometry  study of
the Magellanic  Clouds --  see their  $1.1 < B-R  < 1.35$  diagram. At
least this spiral  arm might be due to tidal action,  as it appears to
be connected  to the Magellanic  Cloud Bridge (cf.   Staveley-Smith et
al.~\cite{ss}). The outernmost contour well matches the carbon stars 
by Kunkel et al.~(\cite{kunk}).

\begin{figure}
\resizebox{7cm}{!}{\includegraphics{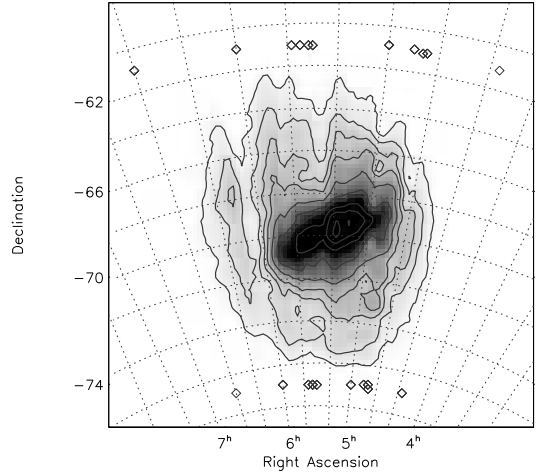}}
\caption{Star count of the LMC -- class C (Fig.~\ref{cml}); contours are at: 100, 150, 200, 250, 350, 500, 600, 800, 1000, 1200 per $0.04$ deg$^2$.}
\label{lmcc}
\end{figure}

The oldest population (from $1$  to $5$ Gyr), represented by RGB stars
(Fig.~\ref{lmcc}),  once  again  reveals  a  prominent  Bar  which  is
significantly   broader    than   that   defined    by   the   younger
populations.  Galactic  foreground  stars  may  affect  the  outermost
contours. The southern spiral arm  is inconspicuous, but the two faint
northern spiral  arms seen in Fig.  3 (AGB) have  weak counterparts in
the form of extensions at $\alpha = 5^h \to 6^h, \delta = -64\degr \to
-68\degr$.

Bothun \& Thompson (~\cite{bt}) conclude that the LMC has a relatively
large scale  length more appropriate for galaxies  with obvious spiral
structure than  for other dwarf  galaxies. It is interesting  that the
asymmetric spiral structure delineated  by the different components in
Figs.~\ref{lmca}--\ref{lmcc} is  in fairly good agreement  with the HI
map shown by Gardiner et al.~(\cite{gard}) and is nicely reproduced by
their dynamical model.

\subsection{Structure of the SMC}
\begin{figure}
\resizebox{7cm}{!}{\includegraphics{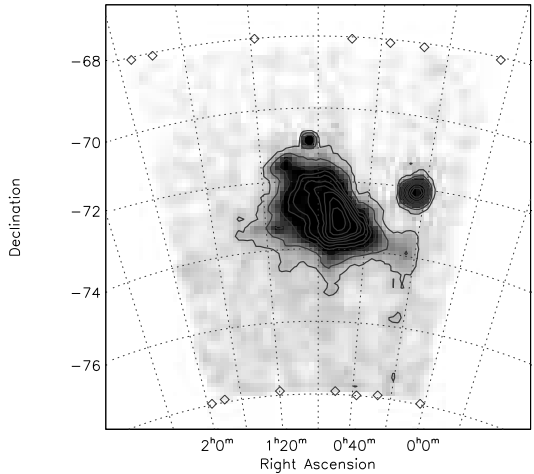}}
\caption{Star count of the SMC -- class A; contours are at: 20, 30, 45, 60, 75, 100, 125, 150, 200, 250 per $0.04$ deg$^2$.}
\label{smca}
\end{figure}

The structure  of the SMC  is still not understood  (Westerlund 1997).
Our  maps  show  that  populations  of different  age  have  different
distributions.    The    youngest   component   has    an   asymmetric
distribution(Fig.~\ref{smca})  elongated  along   a  NE--SW  axis  (PA
$\approx 45\degr$).  In the  south, the outermost contour defines four
protuberances which might be  associated with tidal features: at least
the eastern  (coincident with the SMC Wing)  and western protuberances
are   aligned    with   that   of   the    Magellanic   Cloud   Bridge
(cf.  Staveley-Smith  et  al.~(\cite{ss}).  Higher  contours  show  an
extension in the northeast, aligned with the main body of the SMC Bar.
The Bar structure  itself is similar to that  seen in the distribution
of  young  clusters  (Bica  \&  Dutra  \cite{bd})  and  in  the  upper
main-sequence   map  by   Zaritsky   et  al.~(\cite{zar}).   Clusters,
associations and  HII regions (Bica \& Schmitt  \cite{bisch}) are also
found at the locations of the southern protuberances.  The young stars
are strongly  concentrated in  the southwestern part  of the  SMC Bar.
Outside the main  body of the SMC, the  two Galactic globular clusters
NGC~104 = 47~Tuc  (west) and NGC~362 (north) can  be discerned. The HI
column  density contours  in the  map presented  by  Stanimirovi\`c et
al.~(\cite{stan}) outline  the distribution  of the young  stars quite
well.

\begin{figure}
\resizebox{7cm}{!}{\includegraphics{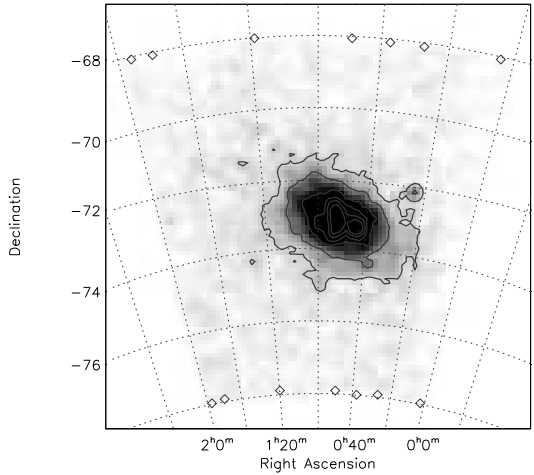}}
\caption{Star count of the SMC -- class B; contours are at: 2, 5, 10, 15, 20, 25, 50 per $0.04$ deg$^2$.}
\label{smcb}
\end{figure}

The AGB stars have  a more regular distribution (Fig.~\ref{smcb}) with
two prominent  central concentrations matching the carbon stars by 
Hardy et al.~(\cite{hardy}). The  easternmost also coincides with
the  peak of  the  young--star distribution.  
The AGB distribution axis  is much
less inclined (PA $\approx 75\degr$) than that of the younger and very
similar to that  of the RGB star distribution.  As in  the case of the
LMC, the  stellar distributions become more regular  and smoother with
increasing  age, also  apparent  in the  $B$  and $V$  band images  by
Zaritsky et al.~(\cite{zar}) and for the outer contour in the carbon 
stars by Kunkel et al.~(\cite{kdi}); carbon stars by Rebeirot et al.
~(\cite{reb}) fill the second level countour (Fig.~\ref{smcb}).

\begin{figure}
\resizebox{7cm}{!}{\includegraphics{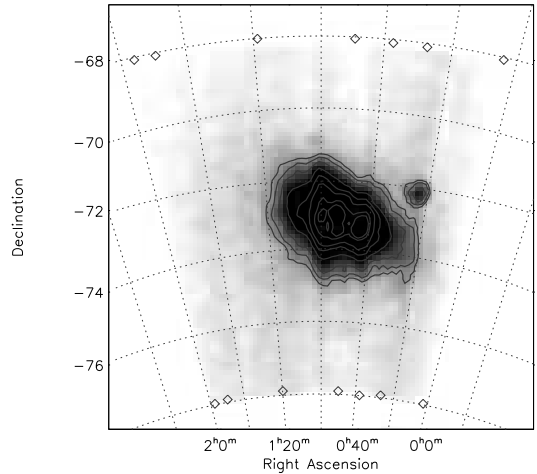}}
\caption{Star count of the SMC -- class C; contours are at: 50, 60, 75, 100, 125, 150, 200, 225 per $0.04$ deg$^2$.}
\label{smcc}
\end{figure}

The distribution of RGB stars  (Fig.~\ref{smcc}) is similar to that of
the  AGB  stars  and  also  exhibits  two  major  concentrations.  The
western most is more  pronounced in RGB than in  AGB stars. The eastern
concentration on average appears  to be significantly younger than the
western concentration  dominated by  the older stars.  Remarkably, the
strongest  HI  concentration  in  the  SMC map  by  Stanimirovi\`c  et
al.~(\cite{stan})  appears to  be  just between  the concentration  of
younger stars and that of older  stars. It is also remarkable that the
older star distribution extends over  the full length of the suspected
southwestern tidal feature,  about $2\degr$ from the main  body of the
Bar. With respect to the overall distribution of the older stars, that
of the  HI appears  to be  displaced towards the  east. The  SMC Wing,
prominent in HI  and also traceable in the  younger stellar population, 
has no counterpart in the older stars.

\section{Conclusions}
Counts  of sources towards  the Magellanic  Clouds extracted  from the
DCMC  (Cioni et  al.~\cite{cio})  allow differentiation  in the  
($I-J$, $I$) colour--magnitude diagram into three groups of objects
with different  mean ages. The  spatial distribution of the  three age
groups is quite different: in either Cloud, the youngest stars exhibit
an irregular  structure characterized  by  spiral  arms and  tidal
features  while  the  older  stars  are smoothly  and  regularly
distributed.   The distribution of  younger stars  is well--correlated
with  those  of  clusters,  associations,  HII  regions  and  HI.  The
significant offset of the LMC Bar with respect to the overall circular
disk suggests that  the LMC potential is dominated  by dark matter. 
The   well--defined  southern  spiral arm may  be   due  to  tidal
interaction with the  SMC. The nature of the  two northern spiral arms
is uncertain.  In the SMC, the regular,  but double--peaked, structure
of the  AGB and  RGB stars is  remarkable, as is its offset from  the HI
distribution,  and  the mean   age  difference of  the  two
maxima.  Relatively  faint east--west  features  in  the younger  star
population  (including  the  Wing)  are  probably also  due  to  tidal
interaction.

\end{document}